\documentclass[letterpaper,conference,final]{IEEEtran}
\IEEEoverridecommandlockouts
\usepackage{hyperref}
\usepackage{amsmath,amssymb,amsfonts}
\usepackage{algorithmic}
\usepackage{graphicx}
\usepackage{textcomp}
\usepackage{biblatex}

\usepackage{array,booktabs}
\newcolumntype{M}[1]{>{\arraybackslash}m{#1}}

\usepackage{xcolor}
\usepackage[ruled,vlined]{algorithm2e}
\usepackage{multirow,multicol}
\usepackage{enumerate}
\usepackage{xcolor}
\usepackage{soul}
\usepackage{tabularx}

\newcommand{\PreserveBackslash}[1]{\let\temp=\\#1\let\\=\temp}
\newcolumntype{C}[1]{>{\PreserveBackslash\centering}p{#1}}
\newcolumntype{R}[1]{>{\PreserveBackslash\raggedleft}p{#1}}
\newcolumntype{L}[1]{>{\PreserveBackslash\raggedright}p{#1}}
\begin{document}

\title{Change Summarization of Diachronic Scholarly Paper Collections by Semantic Evolution Analysis}
\author{
\IEEEauthorblockN{Naman Paharia}
\IEEEauthorblockA{\textit{IIT Kharagpur} \\
\textit{}
Kharagpur, India \\
namanpaharia.27@iitkgp.ac.in}
\and
\IEEEauthorblockN{Muhammad Syafiq Mohd Pozi}
\IEEEauthorblockA{\textit{SOC, Universiti Utara Malaysia}\\
\textit{}
syafiq.pozi@uum.edu.my}
\IEEEauthorblockA{\textit{IIR4.0, Universiti Kebangsaan Malaysia}}
\and
\IEEEauthorblockN{Adam Jatowt}
\textit{University of Innsbruck}\\
Innsbruck, Tirol, Austria \\
adam.jatowt@uibk.ac.at}
\maketitle
\begin{abstract}
The amount of scholarly data has been increasing dramatically over the last years. For newcomers to a particular science domain (e.g., IR, physics, NLP) it is often difficult to spot larger trends and to position the latest research in the context of prior scientific achievements and breakthroughs. Similarly, researchers in the history of science are interested in tools that allow them to analyze and visualize changes in particular scientific domains. Temporal summarization and related methods should be then useful for making sense of large volumes of scientific discourse data aggregated over time.
We demonstrate a novel approach to analyze the collections of research papers published over longer time periods to provide a high level overview of important semantic changes that occurred over the progress of time. Our approach is based on comparing word semantic representations over time and aims to support users in better understanding of large domain-focused archives of scholarly publications. As an example dataset we use the ACL Anthology Reference Corpus that spans from 
1979 to 2015 and contains 22,878 scholarly articles.
\end{abstract}

\begin{IEEEkeywords}
temporal mining, summarization, visualization
\end{IEEEkeywords}

\section{Introduction}

In this paper, the problem we focus on is how to analyze and visualize evolution of scholarly research based on large corpora such as one consisting of computer science scholarly publications, over a period of time. As such datasets typically cover tens of thousands or more articles it is difficult to understand important changes that occurred in the target scientific domain over time. We propose a novel change-oriented summarization approach based on word semantic evolution to provide information on key changes that occurred in the data over time. 
As an underlying data source we use the Association for Computational Linguistics (ACL) dataset, and specifically, the Parcit Structured XML\footnote{https://acl-arc.comp.nus.edu.sg/archives/acl-arc-160301-parscit}
version of the ACL Anthology Reference Corpus~\cite{bird2008acl}. 
By analyzing this data we provide answers to questions about what concepts are highly changing and what are their directions of changes.
A particular research direction that is currently gaining interest in the text summarization field is how to incorporate time factor into automatic text summarization models. 
Examples can been seen in news~\cite{rosin2019generating,yu2021multi,wang2018event}, social networks~\cite{kumar2017temporal}, and scholarly articles~\cite{mohd2020temporal}.
Several strategies have been proposed to understand the significance of individual topics~\cite{yan2015research}, to learn how multidisciplinary research area is founded~\cite{xu2018understanding}, or to illustrate the evolution of scientific communities~\cite{boyack2014creation}. 
A common approach is by frequency analysis~\cite{8791211}, or using topic modelling, so that the correlations between link citations or the sets of co-occurring words are uncovered~\cite{ding2011scientific,dong2018integrated}. 
The problem with those quantitative approaches however, is that, the semantic perspective is usually not considered explicitly during the modelling task. 
Recently a common way of identifying whether a word has changed its meaning over time is by measuring how much its embedding vector has changed through tracking shifts in the word's distributional neighborhood~\cite{kutuzov2018diachronic,soni2020follow,hamilton2016diachronic,tahmasebi2019survey}. 
However, the diachronic word embedding approaches typically focus on individual words aiming to characterize their evolution. On the other hand, in this research, we aim for a collection-level approach which would allow understanding major changes in the entire document collection over time.
\vspace{-.5em}
\section{Method}
\subsection{Data Collection and Preprocessing}
The ACL Anthology Reference Corpus consists of conference and journal papers in natural language processing and computational linguistics research domains which are included within ACL Anthology. The corpus contains 22,878 papers published during a span of 36 years from 1979 to 2015. The distribution of the number of research papers per year is non-uniform, with increasing numbers of papers in the recent years. 
In this work, we use the body text and the year of the publication date of each paper to analyse how specific terminology drifted from 1979 to 2015.

The dataset was divided into 5 non-overlapping time frames: 1979-1995, 1996-2000, 2001-2005, 2006-2010, and 2011-2015. The skewness in data distribution was considered while making this divisions so the time frames have more or less similar number of documents. 
A vocabulary, $V$, was created using the complete dataset after filtering it for stopwords and numbers and extracting nouns and noun phrases. The vocabulary was then filtered to remove words with frequency smaller than 100 as well as to remove words
which occurred only in a single timeframe since we focus on the changes of high significance. Finally, words with length less than 3 were discarded. 

    \vspace{-.5em}
\subsection{Text Representation} 
For representing terms, we used a BERT-based model, called SciBERT~\cite{beltagy2019scibert} with 12 attention layers and the hidden layer of size 768 which was pretrained on 1.14M papers from Semantic Scholar\footnote{\url{https://www.semanticscholar.org/}} resulting in a corpus size of 3.17B tokens.
 SciBERT was then fine-tuned on the entire ACL corpus to represent semantic meanings of words in each time frame.


\vspace{-.5em}
\subsection{Representing Semantic Word Change} 
Analysing topical changes in scientific corpora as well as detecting emerging trends have been traditionally done by frequency analysis or topic modelling. A different approach, which we propose, is to compare semantic meanings of words, using the contextualized embeddings. 
Comparing average embeddings of the same word in different time frames will give a representation of the change of a word. 
In particular, the cosine similarity between the average embedding of a target word in the first time frame (1979-1995) and the average embeddings of this word in the last time frame (2011-2015) will represent the overall semantic change for the word such that the lower the similarity value, the higher the semantic change expected. Tab.~\ref{tab:top20terms} shows words with the lowest and the highest similarity values in our dataset.
\vspace{-1em}
\begin{table}[htbp]
  \centering
  \footnotesize
  
  \caption{Terms with the smallest and highest cosine similarity between the last and first time unit.}
    \begin{tabular}{cc|cc}
    \toprule
     Terms & Smallest Similarity & Terms & Highest Similarity\\
    \midrule
    seventh   &0.6420 & must   &0.9975\\
    fell   &0.6728 &  example   &0.9973\\
    plan   &0.7319 &  words   &0.9972\\
    mann   &0.7648 &  corpus   &0.9970\\
    entail   &0.7849 &  node   &0.9970\\
    west   &0.7982 &  japanese   &0.9969\\
    copyright   &0.7996 &  probability   &0.9969\\
    maxim   &0.8000 &  terminal   &0.9968\\
    prop   &0.8143 &  threshold   &0.9968\\
    jack   &0.8190 &  complex   &0.9966\\
    stanford   &0.8212 &  automaton   &0.9966  \\
    \bottomrule
    \vspace{-1em}
    \end{tabular}%
  \label{tab:top20terms}%
\end{table}%


Table 1 suggests that cosine similarity alone on the level of individual terms does not suffice for summarizing semantic change in the entire corpus. The results for high cosine similarity are rather intuitive, 
but the results corresponding to the low values of similarity tend to deviate from our expectation. These words underwent largest semantic movement yet they seem to carry little significance for our task, neither offer any valid interpretation. Named entities (\texttt{mann}, \texttt{jack}, \texttt{stanford}), numbers (\texttt{seventh}, \texttt{one}), or words like \texttt{copyright}, \texttt{west}, \texttt{program} and \texttt{volume} seem to be coincidentally used in different contexts in different time frames. 

Thus, to recognize and understand key semantic shifts on the level of entire collection, we explore the possibility of grouping words with related meaning drifts.
For each word $ w_i \epsilon V$ we define a difference vector $\vec{d_{w_{i}}}$ representing the difference of its average embedding in the first time frame $\vec{u_{w^{1}_i}}$ and the one in last time frame $\vec{u_{w^{2}_i}}$. 
\begin{equation}\small
  \vec{d_{w_{i}}} =  \vec{u_{w^{1}_i}} - \vec{u_{w^{2}_i}}
\end{equation}
This difference vector captures the information representing the meaning change of a term from the first to the last time frame. 
To better understand this concept, let's take an example of the term \texttt{machine}. Its difference vector will point to words which the term \texttt{machine} moved away from, i.e. mechanical machine, factory and so on, towards the context uses such as algorithm, learning. The representation of this semantic change from 1979 to 2015 will be embedded in the difference vector. 

To explore the drifts of target words we compute the cosine similarity values between the difference vector of $w_i$ ($\vec{d_{w_{i}}}$) and other words in vocabulary ($\vec{u_{w^{1}_j}}$) to find the concepts from where the target word $w_i$ moved away (lowest similarity) and the words that $w_i$ moved towards (highest similarity). In Table~\ref{tab:movement} we show the obtained top words with the lowest and highest similarity values for few selected target words $w_i$. For example, we can observe there that the word \texttt{merit} in 1980s meant the quality of being worthy or praised, yet it drifted towards the concept of evaluation in ML models in 2011-2015. Similarly, the word \texttt{miner} in 1980s represents a person working in mines but it is related to data extraction and information retrieval in the recent time.

\vspace{-1em}
\begin{table}[htbp]
  \centering
  \scriptsize
  \caption{Word movement pattern}
    \begin{tabular}{lp{11.0em}p{12.0em}}
    \toprule
    Terms & \multicolumn{1}{c}{Diverted From} & \multicolumn{1}{c}{Moved To} \\
    \midrule
    \vspace{-0.1cm}
     web &  stein, fin, ray, ink, fell & document, advertising, words, text, citation, sentiment \\\midrule
    \vspace{-0.1cm}
    merit  &  worthy, worth, warrant, wish, sake, permit & confidence, performance, improvement, accuracy, delta, slope\\\midrule
    \vspace{-0.1cm}
    intercept &  attack, kill, shoot, combat, fight & beginning, logarithm, effects, latitude, centroid\\\midrule
    \vspace{-0.1cm}
    machine & workstation, coordinator, factory, constitution, graphic & bayes, learning, reinforcement, radial, regression \\\midrule
    \vspace{-0.1cm}
    miner & work, mining, father, husband & induction, information, extraction, statistics, retrieval \\\midrule
    \vspace{-0.1cm}
    reinforcement  &  justification, proliferation, viability, uniformity  &  machine, descent, bayes, markov, radial\\\midrule
    \vspace{-0.1cm}
    activation  &  elaboration, instantiation, initiation, manifestation & radial, loss, hinge, regularization, gradient\\
    \bottomrule
    \\
    \vspace{-3.9em}
    \end{tabular}%
  \label{tab:movement}%
\end{table}%

\subsection{Change-Oriented Term Grouping}
For summarizing the semantic changes in the entire corpus, we need to analyse the relative movements of multiple words in aggregate. 
As exposed above, some words have meaningful semantic drift, while some are subject to random or no change. Grouping words by the patterns of their semantic drift should help to remove random and meaningless semantic changes (such as some examples in Table~\ref{tab:top20terms}) and will allow to find important changes.
In other words, we look for relations between the movement patterns of different words. This relation can be positive, meaning that it characterizes a group of words that converge in terms of their semantics by moving towards the same point in the semantic vector space, even though at first the words may have no or have only limited shared semantics. Alternatively, the relation can be negative, for the words that divert apart from each other. 

To summarize semantic changes, we cluster words which come from different semantic areas (have different meanings) but which "converge" together over time. We do this by using words' difference vectors (Eq. 1) as the criterion for word similarity. As the number of clusters is not available beforehand and can not be easily estimated, we use Affinity Propagation (AP) Algorithm ~\cite{Frey07clusteringby}.
In AP clustering algorithm one does not need to specify the initial cluster centers as the method automatically finds a subset of exemplar points which can best describe the data groups. 
Since the clustering involves words' difference vectors, the words in a cluster could have different initial semantic meaning yet they all converge together or move towards the same semantic meaning. Example words like \texttt{Architecture}, \texttt{Input}, \texttt{Parameter}, \texttt{Weight}, \texttt{Dimensions} are unrelated to each other and have different meanings in the initial time period (1985-1995) but based on the subset of the scientific corpus spanning 2011-2015, these words became related to deep learning models. 

To further filter the results, we discarded clusters with less than 5 words and removed words within the first quartile of magnitude of their difference vectors ($\vec{d_{w_{i}}}$) to ensure that only words with a significant semantic change are contained in the resulting clusters.
\subsection{Cluster Ranking}
Clusters produced in the above change-oriented clustering represent the temporal movement of words' semantics. The output of Affinity Propagation results in 340 clusters. 
We then need to determine the cluster quality to detect most important and informative clusters.
For this we exploit an observable property of the clusters that, as the words converge together to the similar semantics, they appear together more frequently in the sentences than before. Thus for each cluster, we calculate the frequency counts of the occurrences of this cluster's members in the same sentences in each analyzed time frame. In other words, for each cluster we determine the number of sentences having high containment of the cluster's words in both the first and the last time frame. 
Effectively, we construct the discrete probability distribution of sentences over the number of cluster words occurring in the same sentences in a given time frame.

 To clarify, let \textit{X} be the set of records in \textit{x}, where \textit{x} is the set of all the sentences in a time frame and \textit{Y} be the integer representing the number of words common in a sentence and cluster. Thus for a cluster \textit{i}, the probability of sentences having \textit{n} words in common with the cluster (or $P_i(X \cap y=n)$) will be the number of sentences with \textit{n} common words divided by the total number of sentences in a time frame. Considering the number of cluster members in the same sentence to be utmost 10, the discrete probability distribution for the frequency count of a cluster will be:
\begin{equation}
    y_i = [P_i(X \cap y=1), P_i(X \cap y=2), ... , P_i(X\cap y=10)]
\end{equation}




In particular, Wasserstein Distance (called also Earth Mover's Distance or EMD) is used to calculate the distance between these two distributions. Given two random distributions, EMD can be conceptually portrayed as the task of taking a mass of earth (one distribution) to spread it in space understood as a collection of holes (another distribution) in the same space. EMD measures the least amount of work needed to fill the holes with earth. Thus for a given cluster, the higher the work needed, the higher the semantic importance and novelty of that cluster. Note that EMD is good for our case as there are ordinal relations between units of the distribution. For example, sentences with 4 words from the same cluster are more important than sentences with 3 words which in turn are more important than sentences with 2 words. 
If for a cluster there are many sentences containing large numbers of its words (for example, sentences with 4 or more words from the cluster) it means this cluster is quite coherent. EMD in our case is used for measuring the difference in the cluster's coherence at the last and the first time frames. Suppose there is a cluster for which there were fewer sentences with high number of its member words in the first time unit compared to last, in this case, EMD will be high and the cluster will be judged as important since it represents words that converged to each other over time. 
EMD is defined to minimize the following equation:
\begin{equation}\small
    WORK(U,V,F) = \sum_{i=1}^{m}\sum_{j=1}^{n}d_{i,j}f_{i,j}
    \label{eq:emd}
\end{equation}
where $d_{ij} = d(y_i,y_j)$ is the ground distance between $y_i$ and $y_j$, and $F=[f_{ij}]$ such that $f_{ij}$ is the flow that needs to be determined between $y_i$ and $y_j$. In this work, variables in Eq.~\ref{eq:emd} are:
\begin{table}[htbp]
    \begin{tabularx}{\linewidth}{l@{}c@{}X}
        $y_i$ & $\;:\;$ & discrete probability distribution for $i_{th}$ cluster in $1^{st}$ time frame. \\
        $y_j$ & $\;:\;$ & discrete probability distribution for $j_{th}$ cluster in $5^{th}$ time frame. \\
        $U$ & $\;:\;$ & a matrix of $y_i$ for all i in $1^{st}$ time frame. \\
        $V$ & $\;:\;$ & a matrix of $y_j$ for all j in $5^{th}$ time frame. \\
    \end{tabularx}%
\end{table}%

The top-ranked clusters by this metric are given in Table~\ref{tab:Ranking}.

\begin{table}[htbp]
  \centering
  \footnotesize
  \caption{Top-ranked clusters using Earth Mover's Distance}
    \begin{tabular}{cp{22.445em}}
    \toprule
    $Value$     & \multicolumn{1}{c}{Terms} \\
    \midrule
    \multirow{1}{*}{2317.15}     & data, set, label, split, tuning, subset, class, fold, generalization, validation \\
    \vspace{0.05cm}
    \multirow{1}{*}{2068.34}     &  standard, test, real, deployment, trial \\\vspace{0.05cm}
    \multirow{1}{*}{1737.55}     & model, variable, fit, space, interpolation, bayes, predictor, cue, cache, novelty \\
    \vspace{0.05cm}
    \multirow{1}{*}{1611.03}     & integer, factor, permutation, discount, inflation, multiplicity \\
    \vspace{0.05cm}
    \multirow{1}{*}{1376.89}     & submission, pipeline, evaluation, setting, official, configuration, setup, backup \\
    \vspace{0.05cm}
    \multirow{1}{*}{1374.19}     & result, system, classifier, tie, approach, counterpart, detector \\
    \bottomrule
    \end{tabular}%
  \label{tab:Ranking}%
\end{table}%

As it can be observed from the table, most of the top clusters (cluster 1, 3, 5 and 6) consist of the words which drifted towards a general context of data-science and machine learning, as expected for the computational linguistics corpus.

\vspace{-0.5em}
\section{Experimental Analysis}
We perform a pilot study of the proposed approach using the ACL Anthology corpus and LDA-based baseline.
Latent Dirichlet Allocation (LDA)~\cite{blei2003latent} is a generative probabilistic model for collections of discrete data (e.g., text corpora) with the goal to map all the documents to topics such that the words in each latent topic tend to co-occur with each other. In our experiment we utilize LDA\footnote{\url{https://radimrehurek.com/gensim/models/ldamodel.html}} to monitor the change of topic importance from the time frame 1 to 5 and we compare the results with the ones by the proposed change-oriented clustering. LDA model was trained on the combined dataset of documents from the time frame 1 \& 5, with fine-tuned hyper-parameters such as the number of topics equal to 100 with 20 passes through corpus along with filtering words that occur in less than 30 and more than 75\% of the documents (other parameters had default values). LDA outputs a document-topic matrix representing the importance of the topics per each document. The topics were then sorted in the increasing order of the difference between their importance scores in time frame 1 and ones in
time frame 5. The top 10 topics that gained most importance (i.e., become more dominant) were then selected as the results.

\subsection{Results}
Table~\ref{tab:expresults} shows the comparison between the cluster scores for EMD-based semantic change-oriented clustering which is our proposed model vs. Latent Dirichlet Allocation based ranking. 3 expert reviewers who have worked in NLP field for at least 5 years, were asked to score the top 10 clusters returned by each clustering method on the basis of their quality using 0-5 Likert scale (0 means meaningless results while 5 indicates the highest quality results). The average score from the judges was taken as the final score for each cluster.
Table~\ref{tab:expresults} shows the results for the top 1, 3, 5 and 10 clusters from both the rankings. Change-oriented clustering outperforms LDA baseline in all cases with an overall score improvement of 16\%. Additionally, a consistent trend of the decreasing scores when moving from the top-1 to top-10 signifies the effectiveness of the proposed EMD based ranking. 
\begin{table}[htbp]
  \centering
  \footnotesize
  \caption{Experimental results.}
    \begin{tabular}{ccccc}
    \toprule
    Terms & Top 1 & Top 3 & Top 5 & Top 10 \\
    \midrule
    \vspace{0.1cm}
    EMD & 4.66 & 3.88 & 3.5 & 3.76 \\
    \vspace{0.05cm}
    LDA & 2.66 & 2.33 & 2.77 & 2.96 \\
    \bottomrule
    \vspace{-2em}
    \end{tabular}%
  \label{tab:expresults}%
\end{table}%

\section{Conclusion}
We propose a novel method to identify changing concepts from a large scientific document corpora that are representative for semantic evolution of this corpora. Our approach can help in better understanding scientific document collections and the important changes that are latent in them over time.  
The approach we developed is based on analyzing temporal changes in semantics of terms and on quantifying as well as aggregating their drifting patterns.

In the future, we will find representative sentences for each important cluster to provide a constrastive type summary of the semantic drift underlying the cluster and to better represent each output cluster.
\section*{Acknowledgements}
We are grateful for the support from the Malaysia Ministry of Higher Education\footnote{FRGS/1/2019/ICT02/UUM/02/2} and the free use of Rescale\footnote{https://techagainstcovid.com/} resources. 

\printbibliography
\end{document}